\documentclass{amia}
\usepackage{graphicx}
\usepackage[labelfont=bf]{caption}
\usepackage[superscript,nomove]{cite}
\usepackage[dvipsnames]{xcolor}
\usepackage[sort,comma,numbers,super]{natbib}
\usepackage{multirow}
\usepackage{url}
\usepackage{soul}
\usepackage{hyperref}
\usepackage{listings}
\usepackage{amsmath}
\usepackage{needspace}

\newcommand\EatDot[1]{}

\definecolor{myblue}{HTML}{4E79A7}
\definecolor{mygreen}{HTML}{59A14F}
\definecolor{myred}{HTML}{E15759}

\begin{document}

\title{Patient Cohort Retrieval using Transformer Language Models}

\author{Sarvesh Soni, MS, Kirk Roberts, PhD}

\institutes{
    School of Biomedical Informatics \\
    The University of Texas Health Science Center at Houston \\
    Houston TX, USA \\
}

\maketitle

\noindent{\bf Abstract}

\textit{We apply deep learning-based language models to the task of patient cohort retrieval (CR) with the aim to assess their efficacy. The task of CR requires the extraction of relevant documents from the electronic health records (EHRs) on the basis of a given query. Given the recent advancements in the field of document retrieval, we map the task of CR to a document retrieval task and apply various deep neural models implemented for the general domain tasks. In this paper, we propose a framework for retrieving patient cohorts using neural language models without the need of explicit feature engineering and domain expertise. We find that a majority of our models outperform the BM25 baseline method on various evaluation metrics.}

\section*{Introduction}

Automatic methods for rapidly identifying groups of patients (cohorts) based on a defined set of criteria have numerous biomedical use cases.
Some of these include clinical trial recruitment, survival analysis, and outcome prediction alongside various other retrospective studies \cite{shivade2014ReviewApproachesIdentifying}.
Electronic health records (EHRs) provide a useful means to efficiently extract such cohorts as these clinical records contain large amounts of patient-related information \cite{sarmiento2016ImprovingPatientCohort}.
However, much of the clinically relevant information is present in an unstructured format (e.g., discharge summaries, progress notes) as they allow for rich data entries and more expressiveness.
This also makes it challenging to parse the information present in these notes because of various issues (related to the use of natural language) such as lack of grammar, use of abbreviations, and implicit referrals to other parts of the record \cite{friedman2002TwoBiomedicalSublanguages, voorhees2013TRECMedicalRecords}.

There is a surge of deep learning (DL)-based methods for processing and extracting information from natural language text, both in general and the clinical domain \cite{lecun2015DeepLearning, wu2020DeepLearningClinical}.
These methods are shown to perform better than the traditional machine learning (ML) methods on a variety of natural language processing (NLP) tasks in the open domain \cite{bengio2013RepresentationLearningReview, lecun2015DeepLearning} as well as the clinical domain \cite{wu2020DeepLearningClinical}.
Though numerous work applied DL techniques to clinical information extraction or IE (e.g., named entity recognition, relation extraction) to improve its performance, less focus is drawn toward applying these advanced techniques to information retrieval or IR (e.g., cohort retrieval) from the EHRs \cite{wu2020DeepLearningClinical}.
The fundamental difference between IE and IR is that the former focuses on extracting pre-defined structured information from unstructured text while the latter aims at quickly returning a set of documents relevant to a user-provided, often \textit{ad hoc}, query.

The task of patient cohort retrieval (CR) is to search a large set of EHR documents for determining groups of patients most relevant to a given query.
E.g., one might need to screen \textit{``children with dental caries''} for a clinical trial.
In this scenario, we expect a CR system to return a list of patients who are under the age of $18$ (\textit{children}) and had a problem with \textit{dental caries} at some point in time.
Such type of matching (from query to relevant documents) requires an understanding of the general vocabulary (e.g., \textit{children} would mean searching for patients below $18$ years of age), domain vocabulary (e.g., \textit{dental caries} can also be mentioned in the note as \textit{tooth decay} or \textit{cavities}), as well as reasoning (e.g., understanding the meaning of \textit{with}).
Some of these challenges can be addressed using advanced DL methods \cite{lecun2015DeepLearning, li2018DeepLearningNatural}.

\vspace{-0.95pt}

IR is well researched in the biomedical domain, though much of the earlier work focused on traditional IR approaches (largely non-ML) \cite{hersh2009InformationRetrievalHealth}.
More recently, the patient CR techniques incorporated traditional ML methods \cite{shivade2014ReviewApproachesIdentifying}.
However, the main limitations of these include heavy feature engineering, reliance on domain expertise, and lack of generalizability.
DL techniques aim to overcome some of these limitations by automatically learning the representations from raw data that has the potential to be generalized to other kinds of data \cite{bengio2013RepresentationLearningReview}.
Recently, various language modeling methods have emerged that utilize contextualized representations and are successful in improving the performance of various NLP tasks in both general \cite{devlin2019BERTPretrainingDeep, yang2019XLNetGeneralizedAutoregressive} and specific domains \cite{alsentzer2019PubliclyAvailableClinical, lee2019BioBERTPretrainedBiomedical, huang2019ClinicalXLNetModeling}.
Such language models (LMs) have a great potential to be applied to the task of CR as they can effectively encode the nuances of natural language present in clinical notes and the free-text queries.

In this work, we apply DL-based methods to the task of CR with an aim to improve its performance and generalizability without relying on cumbersome feature engineering and domain expertise.
The unit of retrieval for our task is a visit (i.e., we aim to return a list of relevant visits given a free-text query).
A visit is defined as a single stay of a patient at a hospital \cite{voorhees2013TRECMedicalRecords}.
In other words, all the reports associated with a visit are collected during a single patient encounter and hence these visits can be considered a proxy for retrieving relevant \textit{patients}.
The reason why this proxy evaluation is necessary is described below.
To our knowledge, this is the first work to assess the effectiveness of employing deep language models to the task of patient CR.

\section*{Background}
\setcounter{section}{0}

There was a rise in the implementation of CR systems after the launch of a Medical Records Track (TRECMed) as part of the annual Text REtrieval Conference (TREC) hosted by the National Institute for Standards and Technology (NIST) in $2011$ \cite{voorhees2011OverviewTREC2011}.
Most of the submissions to TRECMed employed a learning-to-rank approach that uses machine learning to rank the resulting documents \cite{goodwin2018LearningRelevanceModels}.
A typical pipeline of the submitted systems included some query expansion component in which an input query is appended with additional terms to increase the model's coverage of results.
Submissions from many track participants employed a concept normalization tool such as MetaMap \cite{aronson2010OverviewMetaMapHistorical} to identify the clinical entities from query text \cite{voorhees2013TRECMedicalRecords, goodwin2018LearningRelevanceModels}.
The identified concepts are used to formulate a machine-understandable query from the textual input.
Some approaches also used disease diagnosis codes associated with the clinical document for query expansion \cite{voorhees2013TRECMedicalRecords}.
Specifically, the descriptions associated with diagnosis codes were used for appending the query.
Additional sources of data for query expansion include synonyms and other related terms \cite{goodwin2018LearningRelevanceModels}.
Moreover, some systems exploited a negation detection system like NegEx \cite{chapman2001SimpleAlgorithmIdentifying} for differentiating between the positive and negative instances of the medical concepts in the text and query \cite{voorhees2013TRECMedicalRecords}.
This was done mainly to exclude the negated terms while searching.
Edinger et al. highlight some of the limitations of the systems submitted to TRECMed \cite{edinger2012BarriersRetrievingPatient}.
Further, approaches based on traditional machine learning are time-consuming and oftentimes require domain expertise to craft a comprehensive set of features \cite{goodwin2018LearningRelevanceModels}.
Finally, such techniques may not be generalizable to other kinds of data.
Thus, we resort to deep learning algorithms that do not require as much extensive feature engineering and domain expertise and are shown to be generalizable to different kinds of data.

\textcolor{black}{
Though the deep learning techniques are not explored as much for the patient CR, it is well-researched in the general domain because of popular use cases such as powering web search engines \cite{mitra2017NeuralModelsInformation, zhu2020ChapterSevenDeep}.
Recently, a plethora of neural models are implemented to rank the documents to efficiently retrieve information from large amount of data \cite{hui2018CoPACRRContextAwareNeural, xiong2017EndtoEndNeuralAdhoc, guo2016DeepRelevanceMatching, nogueira2019PassageRerankingBERT}.
The availability of large IR datasets in the open domain such as TREC-CAR \cite{dietz2017TRECComplexAnswer} and MS MARCO \cite{bajaj2018MSMARCOHuman} have facilitated such advancements as the deep learning techniques usually require large datasets for training.
Further, powerful neural language models (e.g., BERT \cite{devlin2019BERTPretrainingDeep}) have gained a huge popularity among the NLP community due to these models' ability to learn from large unstructured data sources (through pre-training).
Moreover, such models can be fine-tuned on downstream NLP tasks (e.g., document ranking \cite{nogueira2019PassageRerankingBERT}) with minimal tweaks.
We aim to explore the potential of applying such powerful neural language models (in our case BERT) to the task of patient CR.
}

\begin{figure}[t!]
	\centering
	\includegraphics[width=0.8\linewidth]{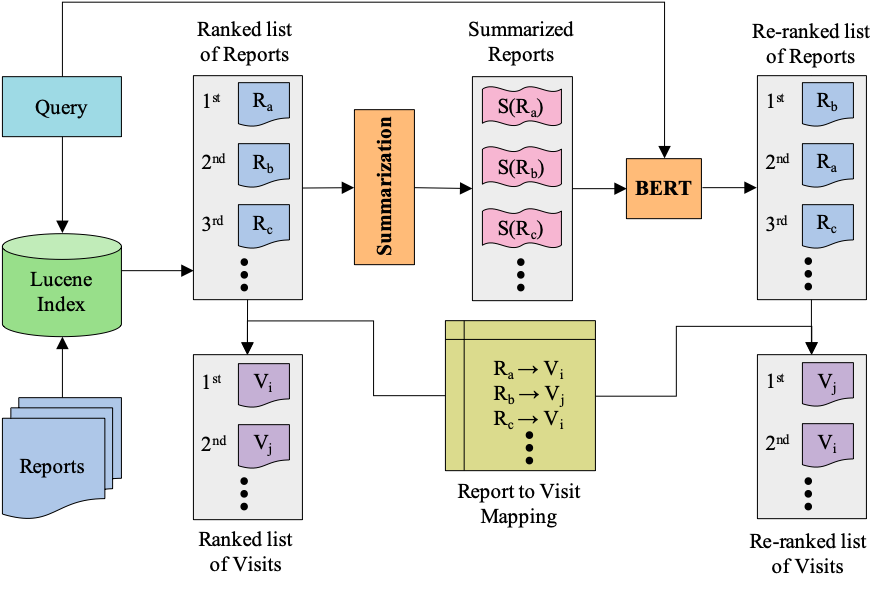}
	\caption{Framework for patient cohort retrieval using BERT. 
	$R_x$ -- Report with identifier $x$, $S(R_x)$ -- Summary for report $R_x$, $V_y$ -- Visit with identifier $y$, $R_x \rightarrow V_y$ -- Report $R_x$ is part of the Visit $V_y$.}
	\label{fig:framework}
\end{figure}

\section*{Materials and Methods}
\setcounter{section}{0}

An overview of the proposed framework is shown in Figure \ref{fig:framework}.
The datasets used in our study are described in Section \ref{data}.
We provide our methods for indexing and querying the large clinical dataset to form a candidate set of documents under Section \ref{candidate}.
The steps for summarizing the information present in EHRs is presented in Section \ref{summarization}.
Further, we explain the re-ranking experiments using BERT in Section \ref{rerank}.
Our evaluation criteria and metrics are described in Section \ref{evaluation} and we provide details about our experimental setup in Section \ref{experiment}.

\section{Data}
\label{data}

The dataset consists of de-identified clinical reports collected from various hospitals during a period of one month.
It was provided to the participants of TREC (Text REtrieval Conference) medical track challenges conducted in $2011$ and $2012$ \cite{voorhees2012OverviewTREC2012, voorhees2013TRECMedicalRecords}.
The corpus contains $93,551$ reports with $17,264$ associated visits.
In particular, each report is linked to a unique visit where a visit can be thought of as a set of patient records related to a single hospital stay.
There are $9$ types of reports in the dataset: Radiology Report, History and Physical, Consultation Report, Emergency Department Report, Progress Note, Discharge Summary, Operative Report, Surgical Pathology Report, and Cardiology Report.

The cohort descriptions (or topics in the terminology of TREC challenges) for both the task years were constructed by physicians.
A total of $34$ and $47$ topics were evaluated as part of the $2011$ and $2012$ tasks, respectively.
The unit of retrieval for this task is visits, i.e., a CR system should return a ranked list of visits corresponding to each topic.
The reason for visit-level evaluations and not patient-level evaluations is that, in order to protect patient privacy, overall patient-level identifiers were stripped from the data and only visit-level identifiers remained.
The original runs submitted by the participants were manually evaluated by physicians.
Each visit from an identified subset of the returned visits by the system was evaluated to be either \textit{relevant}, \textit{partially relevant}, or \textit{not relevant}.
More detailed description of the documents, topics, and relevance judgment procedure is available in the original paper from the task organizers \cite{voorhees2013TRECMedicalRecords}.

\vspace{-3pt}

The relevance judgments for the submitted runs are available to us as part of the dataset and we use these to train and evaluate our pipeline.
We combine the \textit{relevant} and \textit{partially relevant} scores to a single \textsc{relevant} category and keep the \textit{not relevant} scores as \textsc{not relevant}.
Moreover, we assume all the visits not judged for a particular query to be \textsc{not relevant}.

\vspace{-3pt}

\section{Candidate generation}
\label{candidate}

\vspace{-2.5pt}

Though the unit of retrieval for this task is a visit, we index, generate candidates, and rank at the individual report level.
The ranked list of reports are mapped to their corresponding visits in a post hoc manner.
The reason for such an approach is the fact that most of the decisions to mark a visit either \textsc{relevant} or \textsc{not relevant} are based on individual reports and not contingent on the combination of information from multiple reports.
For the scope of our evaluations, we assume that if a visit is labeled \textsc{relevant} or \textsc{not relevant}, the reports included with the visit are also labeled in the same manner.

\vspace{-3pt}

All the reports in the dataset are indexed using Lucene \cite{jakarta2004ApacheLuceneaHighperformance}.
We use Porter stemmer to stem the contents of the reports for indexing.
Stemming ensures that the different inflectional variants of a word can be grouped together.
E.g., all the instances of \textit{dehydrate}, \textit{dehydrated}, and \textit{dehydration} are stored in the index as \textit{dehydr}.
We also leave out stopwords (such as \textit{a}, \textit{an}, \textit{the}) while indexing as they are often not useful for searching.
For this purpose, we use a built-in list of stopwords from Lucene (\lstinline{EnglishAnalyzer} class's \lstinline{ENGLISH_STOP_WORDS_SET}).
For tokenization, we exploit Lucene's \lstinline{StandardTokenizer}.

The cohort description or query is first passed through the same stemming, stopwords filtering, and tokenization steps as was carried out during the indexing process.
The output tokens from these steps are used to construct a bag of words style query for retrieving a ranked set of relevant candidate reports from the constructed index.
E.g., the topic \textit{``Children with dental caries''} is converted to query \textit{``children OR dental OR cari''} that is searched against the index of clinical narratives.
Specifically, we query the index using BM25  \cite{robertson1994OkapiTREC3} as the ranking function.
The ranked list of $N$ candidate reports extracted from the index is passed to subsequent layers for further processing (we use $N=2000$ in our experiments).
We use the publicly available toolkit Anserini to index and query the large set of clinical documents in our dataset \cite{yang2018AnseriniReproducibleRanking}.

\section{Summarization}
\label{summarization}

Before passing the candidate set of reports to our re-ranker model BERT, we summarize their contents.
The motivation behind this is the input length restrictions of the language model we aim to apply.
The sequence length (or length of input) for BERT is determined after wordpiece tokenization that involves dividing a word into sub-word units \cite{wu2016GoogleNeuralMachine}.
A maximum sequence length of $512$ is evaluated by the authors of BERT \cite{devlin2019BERTPretrainingDeep}.
However, the text in clinical reports is much longer than $512$.
We show the descriptive statistics of the number of tokens after wordpiece tokenization of the clinical reports in our dataset in Table \ref{table:token_stats}.

\begin{table}[t]
	\caption{Descriptive statistics of the token counts for TREC Medical Track dataset after wordpiece tokenization. \textbf{SD} -- Standard Deviation. \textbf{Min} -- Minimum. \textbf{Max} -- Maximum.}
	\vspace{-0.15in}
	\centering
	\begin{tabular}{c c c c c c}
		\hline 
		\multirow{2}{*}{\textbf{Component}} & \multicolumn{5}{c}{\textbf{Metric}} \\
		\cline{2-6}
		& \textbf{Mean} & \textbf{Median} & \textbf{SD} & \textbf{Min} & \textbf{Max} \\ 
		\cline{2-6}
		\hline 
		Query & \textcolor{black}{$13.99$} & \textcolor{black}{$13$} & \textcolor{black}{$7.21$} & \textcolor{black}{$4$} & \textcolor{black}{$51$} \\
		\hline
		Report & $1192.58$ & $1126$ & $544.38$ & $94$ & $9934$ \\
		\hline
	\end{tabular} 
	\label{table:token_stats}
\end{table}

There are various approaches to summarize the content present in a clinical narrative.
These largely fall into two categories, extractive (extract salient phrases directly from the narrative itself) and abstractive (summarize using the salient ideas present in the narrative by rephrasing).
As the medical records are often written in a succinct manner, only including the relevant clinical information, we can still end up with a relatively large document even after an extractive summarization where a set of important phrases are selected from the document.
Thus, we take an abstractive approach for summarization. An example of our summarization process is presented in Table \ref{table:summarization}.

\begin{table}[t]
\vspace{0.15in}
\caption{An example of the summarization process shown using an excerpt from an emergency department report. The selected concepts after the filtering process are \textcolor{mygreen}{\textit{\textbf{italicized}}} in the report text. Note that this is not a complete report and hence the list of selected concepts is also incomplete.}
\vspace{-0.1in}
\centering
\fontfamily{ppl}\selectfont
\begin{tabular}{|p{6.3in}|}
\hline
\\
\textbf{Report:} \space \ldots \textcolor{myblue}{\textbf{blood pressure}} 114/65, \textcolor{myblue}{\textbf{pulse oximetry}} 98\% on room \textcolor{myblue}{\textbf{air}}.  The \textcolor{myblue}{\textbf{skin}} is warm and dry. The \textcolor{mygreen}{\textit{\textbf{ears}}} are negative. The \textcolor{myred}{\textbf{pharynx}} is not injected. He has an \textcolor{mygreen}{\textit{\textbf{impacted wisdom tooth}}}.  He has multiple \textcolor{mygreen}{\textit{\textbf{dental}}} \textcolor{mygreen}{\textit{\textbf{caries}}}.  He has no \textcolor{myred}{\textbf{trismus}}.  The \textcolor{myblue}{\textbf{neck}} is supple, shotty \textcolor{mygreen}{\textit{\textbf{adenopathy}}}.  \ldots\\
\\
\textbf{Extracted concepts:} \space \ldots \space blood pressure (+) \space pulse oximetry (+) \space air (+) \space skin (+) \space ears (+) \space pharynx (-) \space impacted wisdom tooth (+) \space dental (+) \space caries (+) \space trismus (-) \space neck (+) \space adenopathy (+) \space \ldots \\
\\
\textbf{Selected concepts:} \space emergency department report; \ldots \space ears; \space impacted wisdom tooth; \space dental; \space caries; \space adenopathy; \space \ldots \\
\\
\hline
\end{tabular}
\label{table:summarization}
\end{table}

We pass the text from reports through a standard pipeline of the Apache cTAKES (clinical Text Analysis and Knowledge Extraction System) \cite{savova2010MayoClinicalText} to detect the underlying medical concepts.
It is an open-source NLP system for extracting information from the clinical text that is based on UIMA (Unstructured Information Management Architecture) framework and used often for clinical IE \cite{wang2018ClinicalInformationExtraction}.
Specifically, we make use of their Default Clinical Pipeline.
We further split the contents of the reports into sentences using Stanford's CoreNLP tool \cite{manning2014StanfordCoreNLPNatural}.
These tokenized sentences are then matched for the presence of negation terms from a pre-defined set \cite{weng2020ClinicalTextSummarization}.
All the sentences in a report that are found to contain these negation terms are considered \textit{negated} sentences.
We align the outputs of cTAKES pipeline and our negated sentence detection pipeline to determine the negative concepts.
Thus, we extract a set of positive and negative concepts through this process.

We pass the positive concepts forward for further processing.
Though the negative concepts also contain useful clinical information, we choose to use the positive concepts as a typical cohort description is aimed toward finding the visits with certain clinical attributes.
Also, this technique of filtering out the negations is known to work well for CR \cite{zhu2013ExploringEvidenceAggregation}.
Even after filtering out the negative concepts, we end up with large sets of medical concepts for the reports in our dataset (in $100s$).

To tackle this, we define a heuristic to further filter the number of concepts while not losing out on important clinical information that can be used to effectively identify the visits.
We calculate the document frequency (the number of reports in which a particular concept is found) for each concept identified in the reports from our corpus.
For each report, we further filter the concepts on the basis of this document frequency.
Specifically, we only keep the concepts that occur in less than $2500$ reports (i.e., with the document frequency of $2500$ or less).
This number is determined to be useful after experimentation.

We also add the report type to the filtered list of concepts. This is done to convey an important aspect of the setting in which the information was recorded. E.g., we add \textit{emergency department report} to the list of concepts extracted from a report that is written for an emergency room encounter.

\section{Re-ranking using BERT}
\label{rerank}

We formulate our task of patient CR as a document retrieval task at the report level.
In other words, we aim to rank a set of documents with respect to a given query.
In our case, the query is the cohort description and the documents are the candidate set of clinical reports.
As explained earlier, we pass a succinct set of relevant clinical concepts for a candidate report instead of passing the whole text.

We use BERT \cite{devlin2019BERTPretrainingDeep} to re-rank the candidate clinical reports.
Specifically, we pass the query along with the summarized content from candidate report to BERT.
The summarized content is formed by combining the relevant clinical concepts extracted from the report.
Using the terminology from the authors of BERT, we pass the query text packed together with the summarized content corresponding to the candidate report.
For instance, a query along with the example report from Table \ref{table:summarization} as a candidate is passed in the form of the following sequence to BERT:

\vspace{1pt}
\lstinline{[CLS] Children with dental caries. [SEP]} \\
\lstinline{emergency department report; impacted wisdom tooth; dental; caries; ... [SEP]}
\vspace{1pt}

The re-ranking using BERT is carried out by training and running the model as a binary classifier \cite{nogueira2019PassageRerankingBERT}.
Each of the above pairs are assigned one of the two labels based on relevance judgments.
If the visit corresponding to the candidate report in the pair is related to the query we mark the input pair as \textsc{relevant}, otherwise \textsc{not relevant}.
Precisely, we pass the final layer vector for \lstinline{[CLS]} token through a neural network to get the probability of the input pair relevancy.
We aim to reduce the standard cross-entropy loss for these probabilities for all the input query-report pairs.
During testing, we calculate the probability of relevancy for each of the test sequences and rank them according to their probability scores (higher the probability of being \textsc{relevant}, better the rank).

It has been shown that BERT fails to perform well in case of imbalanced data \cite{tayyarmadabushi2019CostsensitiveBERTGeneralisable}.
Our dataset is heavily imbalanced, as most of the candidate documents are labeled as \textsc{not relevant}.
Thus, we experiment with varying levels of positive to negative examples ratio in our training dataset.
We determine and use the ratio of positive to negative examples as $1:10$ in all our experiments.
During the training, we also cap the maximum number of candidates to $1650$ while keeping the ratio of positive to negative examples maintained.
As the average number of \textsc{relevant} documents is around $150$, we keep only a maximum of $11 * 150 = 1650$ documents.
E.g., for $150$ \textsc{relevant} documents, the maximum number of \textsc{not relevant} documents is capped at $1500$.

\section{Evaluation}
\label{evaluation}

Since the original task is at the visit level, we map the ranked list of reports from a model to a ranked list of visits (using a mapping between reports and visits).
\textcolor{black}{
Specifically, each report in the ranked list is mapped to its corresponding visit.
As the mapping from reports to visits is many to one, we may end up with duplicate visits in the ranked list of visits.
So, we de-duplicate this ranked list in such a way that only the highest (better) ranked visit is kept in the final ranked list of visits.
}
Due to this many to one mapping, we may end up with fewer ranked visits than reports after the mapping.
Hence, we start with a larger number of reports ($N$) to be able to generate a ranked list of visits ($M$) with the required length.
We use $N=2000$ and $M=1000$ in our experiments.

We measure the performance of our pipeline by assessing the ranked list of visits returned by our pipeline. For comparisons (baseline) we also measure the performance of the ranked list of documents initially retrieved from the Lucene index using BM25.

We use the standard metrics for document retrieval to evaluate the performance of our framework, namely, P@$10$ (Precision for the first 10 documents), rPrec (R-Precision), MAP (Mean Average Precision), Bpref (Binary Preference) \cite{buckley2004RetrievalEvaluationIncomplete}, NDCG (Normalized Discounted Cumulative Gain) \cite{voorhees2005TRECExperimentEvaluation}, and MRR (Mean Reciprocal Rank).
P@$10$ simply calculates the precision for the first $10$ ranked documents by the IR system.
Similarly, rPrec only considers the ground truth relevant documents among the ranked list produced by the system to calculate precision.
Both of the above systems do not take into account the order or rank of the documents in the system output.
To resolve this, MAP measures a system's precision such that the rank of each document in the ranked output contributes to the overall precision.
In particular, the better the rank of a document, the higher its contribution to the overall system precision calculation, and vice-versa.
Differently, Bpref also considers the fact that there may be relevant documents which are not judged as \textsc{relevant} (the other measures assume that unjudged documents are \textsc{not relevant}).
NDCG considers the different levels of relevance judgments during its calculation.
Specifically, it uses a higher relevance score for documents judged relevant than the ones that are judged partially relevant.
MRR calculates the average of the reciprocal ranks (multiplicative inverse of the rank of the first relevant document) across all the queries.

\section{Experimental Setup}
\label{experiment}

We used an uncased variant of BERT\textsubscript{BASE} and BERT\textsubscript{LARGE} models pre-trained and fine-tuned on various general, biomedical, and clinical domain datasets for our evaluations.
We refer to these variants as BERT\textsubscript{BASE} (pre-trained on general domain text) \cite{devlin2019BERTPretrainingDeep}, BioBERT (pre-trained on the general domain and biomedical text) \cite{lee2019BioBERTPretrainedBiomedical}, Clinical BERT (initialized with BioBERT and further pre-trained on clinical notes) \cite{alsentzer2019PubliclyAvailableClinical}, BERT\textsubscript{LARGE} TREC-CAR (pre-trained on general domain text and fine-tuned on TREC-CAR) \cite{nogueira2019PassageRerankingBERT}, and BERT\textsubscript{LARGE} MS MARCO (pre-trained on general domain text and fine-tuned on MS MARCO) \cite{nogueira2019PassageRerankingBERT}.
All these model variants are fine-tuned for both the TRECMed datasets using the hyperparameters described below.
The BERT\textsubscript{BASE} model variant consists of a total of $110$ million parameters with $12$ layers and self-attention heads with a hidden size of $768$ while the BERT\textsubscript{LARGE} model has $340$ million parameters with $24$ layers and a hidden size of $1024$.
We split the datasets into train and test splits at the topic level in the ratio of $80$$:$$20$.
We train the BERT-based model on our training set and evaluated its performance on the testing set.
For our baseline, BM25, we don't need training.
Hence, we directly evaluate its performance for all the topics in our test sets.

For hyperparameters, we use a maximum query length of $64$ as all the topics in our dataset are covered under this length.
Otherwise, we use the recommended set of parameters from the original authors for fine-tuning the model on our dataset.
Namely, the maximum sequence length is $384$ and the learning rate is $ 3 $ x $ 10^{-5} $.
For the models based on BERT\textsubscript{BASE} we use a batch size of $ 24 $ while for the BERT\textsubscript{LARGE} models, it is set to $ 8 $.
We fine-tune all the models for a total of $ 2 $ epochs.
We use a TensorFlow-based implementation of the model and perform all our experiments on an NVidia Tesla V$100$ GPU ($32$G).

\section*{Results}

The evaluation results of our proposed CR framework on the TREC Medical Track datasets from both the years, $2011$ and $2012$, are presented in Tables \ref{table:results_2011} and \ref{table:results_2012}, respectively.
\textcolor{black}{Summarization, capping the number of candidate documents, and maintaining the ratio of relevant (positive) and non-relevant (negative) examples helped in improving the performance of the models over the BERT variants without such restrictions.}

For the $2011$ data, the BERT\textsubscript{LARGE} model fine-tuned on MS MARCO dataset outperformed BM25 on Bpref and P@$1000$ metrics.
Also, this variant and BERT\textsubscript{LARGE} TREC-CAR consistently performed at similar levels of performance as BM25 on all the other metrics.
Among the BERT\textsubscript{BASE} models, the ones further pre-trained on other clinical and biomedical datasets performed better than the vanilla variant on all the measures except P@$1000$.
Also, the models based on BERT\textsubscript{LARGE} achieved higher scores than the models using BERT\textsubscript{BASE}.
Interestingly, the baseline method outperforms almost all the BERT-based models on a majority of the evaluated metrics.

For the $2012$ data, the BERT\textsubscript{LARGE} MS MARCO variant achieved the best scores in terms of most of the performance measures, namely, MAP, rPrec, Bpref, P@$10$, and P@$1000$.
There was a $30.1\%$ increase in the Bpref score by this variant as compared to the baseline score (using BM25).
The vanilla BERT\textsubscript{BASE} model performed the best in terms of NDCG and MRR metrics.
This variant improves upon MRR as compared to BM25 by a total of $0.1914$ points (a $31.7\%$ improvement).
Both BERT\textsubscript{LARGE} MS MARCO and BERT\textsubscript{BASE} models performed consistently better than the baseline on all the evaluated metrics.
Also, the vanilla BERT\textsubscript{BASE} model achieved better scores over all the measures than the other model variants based on BERT\textsubscript{BASE} as well as the BERT\textsubscript{LARGE} TREC-CAR variant.
Similarly, the baseline method surpasses the performances of Clinical BERT, BioBERT, and BERT\textsubscript{LARGE} TREC-CAR on a majority of the measures.
However, we note that all the model variations outperformed the baseline method on Bpref metric except for Clinical BERT.

\begin{table}[t]
	\caption{\textcolor{black}{Results on the $2011$ TREC Medical Track dataset. All the results are on a held-out test split from the dataset.}}
	\vspace{-0.1in}
	\centering
	\begin{tabular}{l c c c c c c c}
		\hline 
		\multirow{2}{*}{\textbf{Variant}} & \multicolumn{7}{c}{\textbf{Metric}} \\
		\cline{2-8}
		& \textbf{MAP} & \textbf{rPrec} & \textbf{Bpref} & \textbf{P@$\mathbf{10}$} & \textbf{P@$\mathbf{1000}$} & \textbf{NDCG} & \textbf{MRR} \\ 
		\cline{2-5}
		\hline 
		BM25 & $0.2595$ & $0.311$ & $0.387$ & $0.4286$ & $0.0523$ & $0.5979$ & $0.6658$ \\
        BERT\textsubscript{BASE} & $0.1537$ & $0.1694$ & $0.3509$ & $0.2$ & $0.0457$ & $0.4327$ & $0.2463$ \\
        Clinical BERT & $0.1707$ & $0.2133$ & $0.3565$ & $0.2714$ & $0.0453$ & $0.4586$ & $0.4142$ \\
        BioBERT & $0.1897$ & $0.222$ & $0.3798$ & $0.2857$  & $0.0451$ & $0.475$ & $0.3997$ \\
        BERT\textsubscript{LARGE} TREC-CAR & $0.2131$ & $0.2528$ & $0.4378$ & $0.3857$ & $0.053$ & $0.5578$ & $0.5095$ \\
        BERT\textsubscript{LARGE} MS MARCO & $0.247$ & $0.2721$ & $\mathbf{0.4627}$ & $0.4143$ & $\mathbf{0.0536}$ & $0.5883$ & $0.6054$ \\
        \hline
	\end{tabular} 
	\label{table:results_2011}
\end{table}

\begin{table}[t]
	\vspace{0.2in}
	\caption{Results on the $2012$ TREC Medical Track dataset. All the results are on a held-out test split from the dataset.}
	\vspace{-0.1in}
	\centering
	\begin{tabular}{l c c c c c c c}
		\hline 
		\multirow{2}{*}{\textbf{Variant}} & \multicolumn{7}{c}{\textbf{Metric}} \\
		\cline{2-8}
		& \textbf{MAP} & \textbf{rPrec} & \textbf{Bpref} & \textbf{P@$\mathbf{10}$} & \textbf{P@$\mathbf{1000}$} & \textbf{NDCG} & \textbf{MRR} \\ 
		\cline{2-5}
		\hline 
		BM25 & $0.2402$ & $0.2584$ & $0.2692$ & $0.4$ & $0.0625$ & $0.5278$ & $0.6036$ \\
        BERT\textsubscript{BASE} & $0.2488$ & $0.2922$ & $0.3255$ & $0.44$ & $0.0633$ & $\mathbf{0.5722}$ & $\mathbf{0.795}$ \\
        Clinical BERT & $0.1686$ & $0.1989$ & $0.23$ & $0.32$ & $0.0614$ & $0.4753$ & $0.5843$ \\
        BioBERT & $0.2263$ & $0.2573$ & $0.302$ & $0.36$ & $0.0614$ & $0.4993$ & $0.6593$ \\
        BERT\textsubscript{LARGE} TREC-CAR & $0.207$ & $0.2779$ & $0.3154$ & $0.37$ & $0.0635$ & $0.5249$ & $0.5679$ \\
        BERT\textsubscript{LARGE} MS MARCO & $\mathbf{0.2539}$ & $\mathbf{0.311}$ & $\mathbf{0.3503}$ & $\mathbf{0.48}$ & $\mathbf{0.0646}$ & $0.565$ & $0.64$ \\
        \hline
	\end{tabular} 
	\label{table:results_2012}
\end{table}

\section*{Discussion}

We perform an array of experiments to analyze the performance improvements of employing deep learning models to the task of patient CR.
We use different variants of the BERT model (both in terms of the model size and the datasets they are trained on) to study their impact on various performance metrics.
To our knowledge, this is the first work to study the benefits of applying deep language models on CR from the EHRs.

We use a large number of metrics for our evaluations as the evaluation measures are application-specific.
Though we focus on the task of CR, similar methods can be employed to other tasks in the same domain.
E.g., such a retrieval pipeline can also be applied to patient-specific EHR search.
In this example, a higher MRR may be preferred.

Though the annotations are available only at the visit level, we trained our pipeline at the report level.
The main reason being the high number of reports for each visit (as much as $415$).
More sophisticated methods or heuristics can be applied to handle the large amounts of reports per visit.
Such methods can aim at summarizing the content at a visit level (like we do at the level of reports).
Further, the effect of summarization can also be explored in detail as it is a well-studied area and offers many different ways to extract the information from the clinical documents \cite{pivovarov2015AutomatedMethodsSummarization, mishra2014TextSummarizationBiomedical}.

The evaluation performances achieved by other systems on the same datasets \cite{goodwin2018LearningRelevanceModels, voorhees2011OverviewTREC2011, voorhees2012OverviewTREC2012, voorhees2013TRECMedicalRecords} may not be directly comparable because of many reasons.
The systems developed as part of the original submissions to the TREC Medical Track challenge were evaluated manually.
Though, as mentioned earlier, we have access to the relevance judgments from the original submissions, note that the set of annotations is not exhaustive or complete.
Hence, we may have a relevant document in our ranked list of visits that might not be judged during the original submissions and thus may not be considered a \textsc{relevant} document during the evaluations.
Some evaluation measures such as Bpref takes into account the absence of relevant documents that may not be available in the ground truth \cite{buckley2004RetrievalEvaluationIncomplete}.

\textcolor{black}{
For the $2012$ data, $60\%$ of our test topics happened to be in the bottom third in terms of their per-topic inferred NDCG scores (infNDCG, the primary evaluation metric for TRECMed $2012$) computed over all the runs submitted to the challenge that year.
Thus, the topics in our test set were predominantly hard.
Also, the language models used in our evaluation are based on transformers that usually take a huge amount of time even for pre-training and fine-tuning.
The average amount of time required for us to run one experiment was about $2$ hours (even on a powerful GPU with $32$GB memory).
Just for running all the final experiments reported in this paper (on a fold), it took around $ 12 * 2 = 24$ hours.
Thus, performing cross-validation for all the included variants would have been highly resource-intensive.
Besides, the aim of this paper is to highlight the importance of applying transformer language models to the task of patient CR and showing how it can be used to improve the performance over baseline methods such as BM25.
Best practices to incorporate such powerful models to the tasks of specific domains will require more thorough experimentation.
}

\textcolor{black}{
The baseline method outperformed almost all the models variants for the $2011$ data.
However, for the $2012$ data, some model variations were able to consistently outscore the baseline method.
We note that the total number of topics for the $2011$ dataset is less than that for the $2012$ data.
Thus, the models for $2012$ data were trained on a larger training set ($80\%$ of that year's dataset) than what was used to train the $2011$ models.
The larger training set (with $10$ topics more worth of training data) may have played a role in enabling some of the transformer models to outperform the baseline for $2012$ dataset.
}

Unfortunately, the TRECMed dataset is no longer available to general public because of issues related to sharing the clinical data.
This is an ongoing problem in clinical research.
However, domain adaptation and transfer learning techniques can play an important role in overcoming some of the issues related to the shorter or uncleaned supply of data for tuning the models \cite{soni2020EvaluationDatasetSelection}.
E.g., in this work we make use of the BERT models pre-trained on large general-domain data and also employ models fine-tuned on open-domain document retrieval datasets.
We note that the models fine-tuned on open-domain datasets performed consistently better than the baseline and other vanilla models.
This highlights the importance of transferring knowledge from large open-domain datasets for a similar task.

\section*{Conclusion}

We experiment with different variants of transformer language models based on the BERT architecture.
We find that incorporating a language model pre-trained and fine-tuned on various general domain and clinical domain datasets outperforms the baseline BM25 method on a majority of metrics.
Our models achieve up to $31.7\%$ improvement over the performance of the baseline model on several metrics.
To our knowledge, this is the first work to study the impact of employing deep language models to the task of patient CR.

\section*{Acknowledgments}
This work was supported by the U.S. National Library of Medicine, National Institutes of Health (NIH), under award R00LM012104, and the Cancer Prevention and Research Institute of Texas (CPRIT), under award RP170668.

\bibliographystyle{vancouver}
\setlength{\bibsep}{5.5pt}
\setlength{\itemsep}{0.9pt}
\bibliography{amia2020}

\begin{thebibliography}{10}

\bibitem{shivade2014ReviewApproachesIdentifying}
Shivade C, Raghavan P, {Fosler-Lussier} E, Embi PJ, Elhadad N, Johnson SB,
  et~al.
\newblock A Review of Approaches to Identifying Patient Phenotype Cohorts Using
  Electronic Health Records.
\newblock Journal of the American Medical Informatics Association.
  2014;21:221--230.

\bibitem{sarmiento2016ImprovingPatientCohort}
Sarmiento RF, Dernoncourt F.
\newblock Improving {{Patient Cohort Identification Using Natural Language
  Processing}}.
\newblock In: {MIT Critical Data}, editor. Secondary {{Analysis}} of
  {{Electronic Health Records}}; 2016. p. 405--417.

\bibitem{friedman2002TwoBiomedicalSublanguages}
Friedman C, Kra P, Rzhetsky A.
\newblock Two Biomedical Sublanguages: A Description Based on the Theories of
  {{Zellig Harris}}.
\newblock Journal of Biomedical Informatics. 2002;35:222--235.

\bibitem{voorhees2013TRECMedicalRecords}
Voorhees EM.
\newblock The {{TREC Medical Records Track}}.
\newblock In: Proceedings of the {{International Conference}} on
  {{Bioinformatics}}, {{Computational Biology}} and {{Biomedical Informatics}};
  2013. p. 239--246.

\bibitem{lecun2015DeepLearning}
LeCun Y, Bengio Y, Hinton G.
\newblock Deep Learning.
\newblock Nature. 2015;521:436--444.

\bibitem{wu2020DeepLearningClinical}
Wu S, Roberts K, Datta S, Du J, Ji Z, Si Y, et~al.
\newblock Deep Learning in Clinical Natural Language Processing: A Methodical
  Review.
\newblock Journal of the American Medical Informatics Association.
  2020;27:457--470.

\bibitem{bengio2013RepresentationLearningReview}
Bengio Y, Courville A, Vincent P.
\newblock Representation {{Learning}}: {{A Review}} and {{New Perspectives}}.
\newblock IEEE Transactions on Pattern Analysis and Machine Intelligence.
  2013;35:1798--1828.

\bibitem{li2018DeepLearningNatural}
Li H.
\newblock Deep Learning for Natural Language Processing: Advantages and
  Challenges.
\newblock National Science Review. 2018;5:24--26.

\bibitem{hersh2009InformationRetrievalHealth}
Hersh W.
\newblock Information {{Retrieval}}: {{A Health}} and {{Biomedical
  Perspective}}.
\newblock 3rd ed.; 2009.

\bibitem{devlin2019BERTPretrainingDeep}
Devlin J, Chang MW, Lee K, Toutanova K.
\newblock {{BERT}}: {{Pre}}-Training of {{Deep Bidirectional Transformers}} for
  {{Language Understanding}}.
\newblock In: Proceedings of the {{North American Chapter}} of the
  {{Association}} for {{Computational Linguistics}}: {{Human Language
  Technologies}}; 2019. p. 4171--4186.

\bibitem{yang2019XLNetGeneralizedAutoregressive}
Yang Z, Dai Z, Yang Y, Carbonell J, Salakhutdinov R, Le QV.
\newblock {{XLNet}}: {{Generalized Autoregressive Pretraining}} for {{Language
  Understanding}}.
\newblock arXiv. 2019;1906.08237 [cs].

\bibitem{alsentzer2019PubliclyAvailableClinical}
Alsentzer E, Murphy J, Boag W, Weng WH, Jindi D, Naumann T, et~al.
\newblock Publicly {{Available Clinical BERT Embeddings}}.
\newblock In: Proceedings of the 2nd {{Clinical Natural Language Processing
  Workshop}}; 2019. p. 72--78.

\bibitem{lee2019BioBERTPretrainedBiomedical}
Lee J, Yoon W, Kim S, Kim D, Kim S, So CH, et~al.
\newblock {{BioBERT}}: A Pre-Trained Biomedical Language Representation Model
  for Biomedical Text Mining.
\newblock Bioinformatics. 2019;p. 1--7.

\bibitem{huang2019ClinicalXLNetModeling}
Huang K, Singh A, Chen S, Moseley ET, Deng Cy, George N, et~al.
\newblock Clinical {{XLNet}}: {{Modeling Sequential Clinical Notes}} and
  {{Predicting Prolonged Mechanical Ventilation}}.
\newblock arXiv. 2019;1912.11975 [cs].

\bibitem{voorhees2011OverviewTREC2011}
Voorhees EM, Tong R.
\newblock Overview of the {{TREC}} 2011 {{Medical Records Track}}.
\newblock In: The {{Twentieth Text REtrieval Conference Proceedings}}; 2011.
  \EatDot.

\bibitem{goodwin2018LearningRelevanceModels}
Goodwin TR, Harabagiu SM.
\newblock Learning Relevance Models for Patient Cohort Retrieval.
\newblock JAMIA Open. 2018;1:265--275.

\bibitem{aronson2010OverviewMetaMapHistorical}
Aronson AR, Lang FM.
\newblock An Overview of {{MetaMap}}: Historical Perspective and Recent
  Advances.
\newblock Journal of the American Medical Informatics Association.
  2010;17:229--236.

\bibitem{chapman2001SimpleAlgorithmIdentifying}
Chapman WW, Bridewell W, Hanbury P, Cooper GF, Buchanan BG.
\newblock A {{Simple Algorithm}} for {{Identifying Negated Findings}} and
  {{Diseases}} in {{Discharge Summaries}}.
\newblock Journal of Biomedical Informatics. 2001;34:301--310.

\bibitem{edinger2012BarriersRetrievingPatient}
Edinger T, Cohen AM, Bedrick S, Ambert K, Hersh W.
\newblock Barriers to {{Retrieving Patient Information}} from {{Electronic
  Health Record Data}}: {{Failure Analysis}} from the {{TREC Medical Records
  Track}}.
\newblock AMIA Annual Symposium Proceedings. 2012;2012:180--188.

\bibitem{mitra2017NeuralModelsInformation}
Mitra B, Craswell N.
\newblock Neural {{Models}} for {{Information Retrieval}}.
\newblock arXiv. 2017;1705.01509 [cs].

\bibitem{zhu2020ChapterSevenDeep}
Zhu R, Tu X, Xiangji~Huang J.
\newblock Chapter Seven - {{Deep}} Learning on Information Retrieval and Its
  Applications.
\newblock In: Das H, Pradhan C, Dey N, editors. Deep {{Learning}} for {{Data
  Analytics}}; 2020. p. 125--153.

\bibitem{hui2018CoPACRRContextAwareNeural}
Hui K, Yates A, Berberich K, {de Melo} G.
\newblock Co-{{PACRR}}: {{A Context}}-{{Aware Neural IR Model}} for {{Ad}}-Hoc
  {{Retrieval}}.
\newblock In: Proceedings of the 11th {{ACM International Conference}} on {{Web
  Search}} and {{Data Mining}}; 2018. p. 279--287.

\bibitem{xiong2017EndtoEndNeuralAdhoc}
Xiong C, Dai Z, Callan J, Liu Z, Power R.
\newblock End-to-{{End Neural Ad}}-Hoc {{Ranking}} with {{Kernel Pooling}}.
\newblock In: Proceedings of the 40th {{International ACM SIGIR Conference}} on
  {{Research}} and {{Development}} in {{Information Retrieval}}; 2017. p.
  55--64.

\bibitem{guo2016DeepRelevanceMatching}
Guo J, Fan Y, Ai Q, Croft WB.
\newblock A {{Deep Relevance Matching Model}} for {{Ad}}-Hoc {{Retrieval}}.
\newblock In: Proceedings of the 25th {{ACM International}} on {{Conference}}
  on {{Information}} and {{Knowledge Management}}; 2016. p. 55--64.

\bibitem{nogueira2019PassageRerankingBERT}
Nogueira R, Cho K.
\newblock Passage {{Re}}-Ranking with {{BERT}}.
\newblock arXiv. 2019;1901.04085 [cs].

\bibitem{dietz2017TRECComplexAnswer}
Dietz L, Verma M, Radlinski F, Craswell N.
\newblock {{TREC Complex Answer Retrieval Overview}}.
\newblock In: {{TREC}}; 2017. \EatDot.

\bibitem{bajaj2018MSMARCOHuman}
Bajaj P, Campos D, Craswell N, Deng L, Gao J, Liu X, et~al.
\newblock {{MS MARCO}}: {{A Human Generated MAchine Reading COmprehension
  Dataset}}.
\newblock arXiv. 2018;1611.09268 [cs].

\bibitem{voorhees2012OverviewTREC2012}
Voorhees EM, Hersh W.
\newblock Overview of the {{TREC}} 2012 {{Medical Records Track}}.
\newblock In: The {{Twenty}}-{{First Text REtrieval Conference Proceedings}};
  2012. \EatDot.

\bibitem{jakarta2004ApacheLuceneaHighperformance}
Jakarta A. Apache {{Lucene}}-a High-Performance, Full-Featured Text Search
  Engine Library; 2004.
\newblock Apache Lucene.

\bibitem{robertson1994OkapiTREC3}
Robertson SE, Walker S, Jones S, {Hancock-Beaulieu} M, Gatford M.
\newblock Okapi at {{TREC}}-3.
\newblock In: Proceedings of the 3rd {{Text REtrieval Conference}}; 1994. p.
  109--126.

\bibitem{yang2018AnseriniReproducibleRanking}
Yang P, Fang H, Lin J.
\newblock Anserini: {{Reproducible Ranking Baselines Using Lucene}}.
\newblock Journal of Data and Information Quality. 2018;10:16:1--16:20.

\bibitem{wu2016GoogleNeuralMachine}
Wu Y, Schuster M, Chen Z, Le QV, Norouzi M, Macherey W, et~al.
\newblock Google's {{Neural Machine Translation System}}: {{Bridging}} the
  {{Gap}} between {{Human}} and {{Machine Translation}}.
\newblock arXiv. 2016;1609.08144 [cs].

\bibitem{savova2010MayoClinicalText}
Savova GK, Masanz JJ, Ogren PV, Zheng J, Sohn S, {Kipper-Schuler} KC, et~al.
\newblock Mayo Clinical {{Text Analysis}} and {{Knowledge Extraction System}}
  ({{cTAKES}}): Architecture, Component Evaluation and Applications.
\newblock Journal of the American Medical Informatics Association.
  2010;17:507--513.

\bibitem{wang2018ClinicalInformationExtraction}
Wang Y, Wang L, {Rastegar-Mojarad} M, Moon S, Shen F, Afzal N, et~al.
\newblock Clinical Information Extraction Applications: {{A}} Literature
  Review.
\newblock Journal of Biomedical Informatics. 2018;77:34--49.

\bibitem{manning2014StanfordCoreNLPNatural}
Manning C, Surdeanu M, Bauer J, Finkel J, Bethard S, McClosky D.
\newblock The {{Stanford CoreNLP}} Natural Language Processing Toolkit.
\newblock In: Proceedings of 52nd Annual Meeting of the Association for
  Computational Linguistics: {{System}} Demonstrations; 2014. p. 55--60.

\bibitem{weng2020ClinicalTextSummarization}
Weng WH, Chung YA, Tong S.
\newblock Clinical {{Text Summarization}} with {{Syntax}}-{{Based Negation}}
  and {{Semantic Concept Identification}}.
\newblock arXiv. 2020;2003.00353 [cs].

\bibitem{zhu2013ExploringEvidenceAggregation}
Zhu D, Carterette B.
\newblock Exploring Evidence Aggregation Methods and External Expansion Sources
  for Medical Record Search.
\newblock In: Proceedings of the 21th {{Text REtrieval}} Conference ({{TREC}});
  2013. \EatDot.

\bibitem{tayyarmadabushi2019CostsensitiveBERTGeneralisable}
Tayyar~Madabushi H, Kochkina E, Castelle M.
\newblock Cost-Sensitive {{BERT}} for Generalisable Sentence Classification on
  Imbalanced Data.
\newblock In: Proceedings of the Second Workshop on Natural Language Processing
  for Internet Freedom: {{Censorship}}, Disinformation, and Propaganda; 2019.
  p. 125--134.

\bibitem{buckley2004RetrievalEvaluationIncomplete}
Buckley C, Voorhees EM.
\newblock Retrieval Evaluation with Incomplete Information.
\newblock In: Proceedings of the 27th Annual International {{ACM SIGIR}}
  Conference on {{Research}} and Development in Information Retrieval; 2004. p.
  25--32.

\bibitem{voorhees2005TRECExperimentEvaluation}
Voorhees EM, Harman DK.
\newblock {{TREC}}: {{Experiment}} and Evaluation in Information Retrieval.
  vol.~63; 2005.

\bibitem{pivovarov2015AutomatedMethodsSummarization}
Pivovarov R, Elhadad N.
\newblock Automated Methods for the Summarization of Electronic Health Records.
\newblock Journal of the American Medical Informatics Association.
  2015;22:938--947.

\bibitem{mishra2014TextSummarizationBiomedical}
Mishra R, Bian J, Fiszman M, Weir CR, Jonnalagadda S, Mostafa J, et~al.
\newblock Text Summarization in the Biomedical Domain: {{A}} Systematic Review
  of Recent Research.
\newblock Journal of Biomedical Informatics. 2014;52:457--467.

\bibitem{soni2020EvaluationDatasetSelection}
Soni S, Roberts K.
\newblock Evaluation of {{Dataset Selection}} for {{Pre}}-{{Training}} and
  {{Fine}}-{{Tuning Transformer Language Models}} for {{Clinical Question
  Answering}}.
\newblock In: Proceedings of the 12th {{International Conference}} on
  {{Language Resources}} and {{Evaluation}}; 2020. p. 5534--5540.

\end{thebibliography}

\end{document}